\documentclass{article}
\usepackage{amssymb,amsmath,graphicx,multicol}
\usepackage{textcomp}
\usepackage{float}
\usepackage{fixltx2e}
\usepackage{color}
\usepackage{array}
\usepackage{authblk}
\begin{document}
\title{CuCoTrack: Cuckoo Filter Based Connection Tracking}

\author[1]{Pedro Reviriego}
\author[2]{Salvatore Pontarelli}
\author[3]{Gil Levy}
\affil[1]{Universidad Antonio de Nebrija, C/ Pirineos, 55, E-28040 Madrid, Spain. \texttt{previrie@nebrija.es}}
\affil[2]{Consorzio Nazionale Interuniversitario per le Telecomunicazioni (CNIT), Via del Politecnico 1, 00133 Rome, Italy. \texttt{salvatore.pontarelli@uniroma2.it}}
\affil[3]{Mellanox Technologies, Ltd. Hakidma 26, Ofer Industrial Park, Yokneam, Israel. \texttt{gill@mellanox.com}}

\maketitle

\begin{abstract}
This paper introduces CuCoTrack, a cuckoo hash based data structure designed to efficiently implement connection tracking. The proposed scheme exploits the fact that queries always match one existing connection to compress the 5-tuple that identifies the connection. This reduces significantly the amount of memory needed to store the connections and also the memory bandwidth needed for lookups. CuCoTrack uses a dynamic fingerprint to avoid collisions thus ensuring that queries are completed in at most two memory accesses and facilitating a hardware implementation. The proposed scheme has been analyzed theoretically and validated by simulation. The results show that using 16 bits for the fingerprint is enough to avoid collisions in practical configurations.   


\end{abstract}

\section{Introduction}

In many networking applications, there is a need to track each of the TCP connections on a link. This is needed for example for load balancing of servers in datacenter networks \cite{SilkRoad}.  In those applications, each incoming packet has to be matched against the set of existing connections $S$ to retrieve a value that is used to determine the processing of the packet. New connections are identified when the packet has the SYN flag set and then are inserted in the set of active connections.  


The lookup for each packet can be done using cuckoo hashing to store the set of elements $S$ each with its associated value \cite{CH}. With cuckoo hashing, a query for element $x$ to retrieve its associated value $v_x$ can be completed in a small and constant number of memory accesses and close to full memory occupancy can be achieved \cite{CHocc}. Elements can also be added or removed from the set dynamically. This makes it attractive for high speed hardware implementations \cite{Barefoot}, \cite{pipeline}. 

For connection tracking, the size of the key used to identify the connection is large as it is formed by the source and destination IP addresses and ports and the protocol field and it is commonly referred to as 5-tuple. For example, an IPv6 5-tuple has 296 bits and for IPv4 104 bits. This means that operations require large memory bandwidth and also that the size of the memory needed to store the connections in $S$ is significant \cite{SilkRoad}. This poses a limitation to the number of connections that can be stored inside a switch ASIC and processed at high speed. A much larger number of keys could be stored in external memory but then, the access time will be much larger and the number of lookups that can be done per second would be much smaller. 

A key observation, is that for connection tracking, only lookup operations for elements (5-tuples) that are in $S$ are performed (new connections are identified using the SYN field). Therefore, one possibility is to try to compress the elements in $S$ before storing them into the cuckoo hash tables. One way to implement the compression is to store a fingerprint of the element instead the element itself. This fingerprint can be computed using a hash function on the element $f_x = h_f(x)$. This approach has been used for example in \cite{SilkRoad} with fingerprints of 16 bits. In fact, a cuckoo filter \cite{CF} could be potentially used as a compressed cuckoo hash. The problem, is that we need to ensure that when searching for an element $x$, there is no fingerprint for another element $y$ on the buckets assigned to $x$ that has the same fingerprint. This will be referred to as a collision of the elements in the rest of the paper. If that occurs, we could not tell which fingerprint corresponds to $x$ and which to $y$. In \cite{SilkRoad}, this issue was solved by handling collisions with additional lookup stages something that requires additional data structures and processing. 

Collisions can be avoided by using more sophisticated data structures such as the Bloomier filter \cite{Bloomier1},\cite{Bloomier2}. A Bloomier filter can store a value $v_x$ associated with each element so that the correct value is retrieved for all elements in the set, i.e. there are no collisions among elements. For elements not in the set it can retrieve a value (false positive) with low probability. The problem is that Bloomier filters are not easily amenable to hardware implementations and supporting dynamic insertions and deletions is also challenging. For connection tracking, we need to support frequent insertions and removals of elements as connections are by nature dynamic. We also want the scheme to be implementable in switching ASICs. Therefore, the Bloomier filter does not fit with our needs.  

In this paper we present CuCoTrack, a data structure that avoids collisions while supporting dynamic insertions and removals. The proposed scheme is based on cuckoo hashing and thus amenable to hardware implementation. To avoid collisions, CuCoTrack uses an adaptive fingerprint. Those were first introduced in the Adaptive Cuckoo Filter (ACF) to reduce the false positive rate \cite{ACF}. In CuCoTrack, fingerprints have a both fixed part and an adaptive part to ensure that collisions can be avoided.

The rest of the paper is structured as follows, CucoTrack is introduced in section 2 and in section 3 is analyzed theoretically. In section 4, simulation results are presented to show that collisions can be avoided for practical configurations and also to validate the theoretical analysis. Finally the conclusions are summarized in section 5. 

\section{CuCoTrack}

\begin{figure}[th]
  \centering
  \includegraphics[scale=0.70]{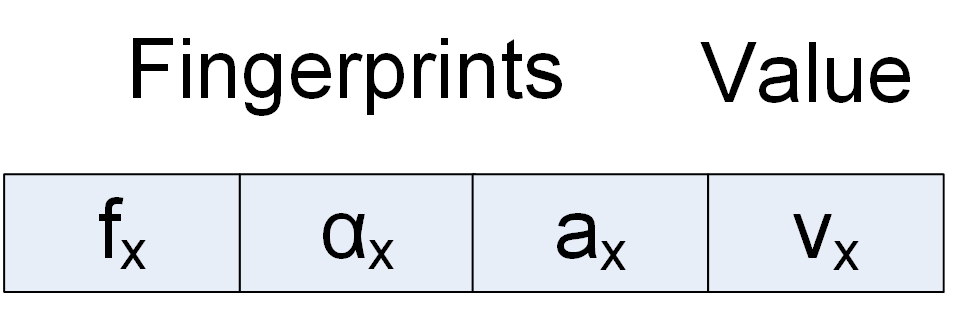}\\
  \caption{Information stored for element $x$ on the proposed data structure.}\label{Fig1}
\end{figure}

\begin{figure}[h!]
  \centering
  \includegraphics[scale=0.7]{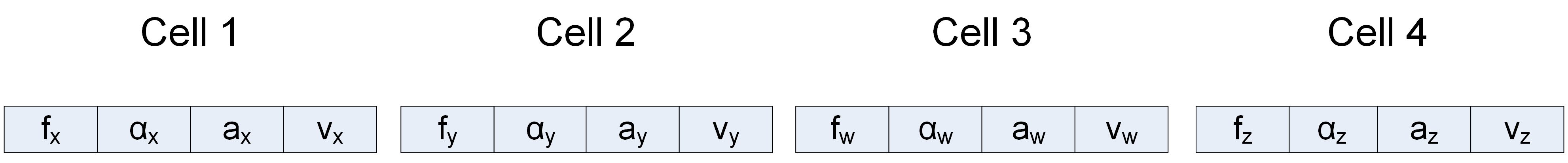}\\
  \caption{Example of a bucket in the proposed CuCoTrack scheme.}\label{Fig2}
\end{figure}

To avoid collisions, the proposed data structure uses a fingerprint that has a both a fixed part and an adaptive part. The fingerprint is defined as follows:

\begin{itemize}
\item Fixed fingerprint: $f_x$ = $h_f(x)$.
\item Adaptive fingerprint composed of: 
\begin{itemize}
\item Hash function selector: $\alpha_x$.
\item Value: $a_x$ = $h_\alpha(x)$.
\end{itemize}
\end{itemize}

where the adaptive fingerprint is similar to one of the schemes proposed for the Adaptive Cuckoo Filters (ACF) \cite{ACF}. The information stored for an element $x$ on the proposed data structure is illustrated in Figure \ref{Fig1} and 
is composed by the fixed and adaptive fingerprints plus the value associated with the element $v_x$. The CuCoTrack structure uses buckets of $c$ = 4 cells as in the original cuckoo filter \cite{CF}. The structure of a bucket composed of four cells is shown in Figure \ref{Fig2}.

Instead of using a single table to which each element $x$ maps to two positions as in the cuckoo filter, CuCoTrack uses $d = 2$ independent tables and each element maps to a single position on each table. In more detail, an element $x$ is mapped to two buckets: $p_1 = h_1(x)$ on the first table and $p_2 = h_1(x)\; xor \; h_2(f_x)$ on the second table. It must be noted that those positions depend only on $x$ and the fixed fingerprint $f_x$. The adaptive part of the fingerprint is not used to compute the positions. This will be one of  the key aspects of the design as will be seen in the following. 

To insert a new element $x$ on the set, we just need to access the two buckets $p_1 = h_1(x)$ on the first table and $p_2 = h_1(x) \; xor \; h_2(f_x)$ on the second table and check that there are no elements stored there that match both the fixed $f_x$ and adaptive $a_x$ fingerprints. If that occurs there is a collision. To avoid it,  we can change the value of the hash selector function $\alpha_x$ and update the adaptive fingerprint $a_x$ = $h_\alpha(x)$ accordingly. To do so, we need to store a copy of the original elements, possibly in a larger and slower external memory (as done in \cite{ACF}).  
The change in $\alpha_x$ has to be done such that the value selected for each colliding element gives a fingerprint that is unique among them. For example, if we have three colliding elements $x,y,z$ and we have four values of $\alpha_x$ so that the values of the hash functions are:

{\centering
 \medskip
\begin{tabular}{ c | c c c}
 $\alpha_x$ & $x$ & $y$ & $z$  \\
\hline         
0 & a & b & b \\

1 & c & a & c \\

2 & a & a & b \\

3 & a & a & a  \\
\end{tabular}

\medskip
}
we can select $\alpha_x$ = 0 for $x$, $\alpha_y$ = 1 for $y$ and $\alpha_z$ = 2 for $z$  and there will be no collisions when we search for any of the elements. We fail to adapt only when there is at least one conflicting element for which all values of $\alpha_x$ give a hash value that is equal to the one of another of the conflicting elements.   


Once collisions are avoided, the insertions proceed as in a standard cuckoo filter \cite{CF} displacing stored fingerprints if needed to make room for the newly inserted fingerprint.
 
The full elements can be stored using tables that have a one to one correspondence of buckets and cells with the CuCoTrack structure.  Then when changing $\alpha_x$, we access the same cell on the full table to compute the new $a_x = h_\alpha(x)$. The fixed cuckoo filter fingerprint $f_x$ is needed to ensure that the bucket locations of an element do not depend on the adaptation. If that were the case, an adaptation could create further collisions with other elements and queries would need to check the positions that correspond to all the values of $\alpha_x$. In our design, cuckoo movements during insertions cannot create new collisions as the movements do not change the locations where an element $x$ can be stored and a new element $y$ that collides with $x$ would map to the same positions as in that case $f_x = f_y$. Therefore, the check for collisions is only needed for new elements added to the set not for movements during insertions. Finally, queries and removals of elements are the same as in a standard cuckoo filter. 

\section{Analysis}

Let us denote as $f$ the number of bits allocated for the fixed fingerprint, $a$ for the adaptive fingerprint value and $\alpha$ for the hash selector. To estimate the probability of the structure failing to avoid collisions, let us consider that the filter has an occupancy of $o$. The analysis can be done by looking at each pair of values given by {$p_1 = h_1(x), f_x$} and {$p_2 = h_1(x)\; xor \; h_2(fx), f_x$}. That is, once we select a fixed fingerprint and a position on the first table, the position on the second table is also determined. Those two positions are the ones that $x$ will map to with fixed fingerprint $f_x$. Therefore only elements that map to those positions with that fingerprint can create a collision with $x$. For two tables with $m/2$ buckets each, there are $m/2 \cdot 2^{f}$ possible pairs of values.  The mapping of the $m \cdot c \cdot o$ elements to the pairs of values can be approximated by a Poisson distribution with $ \lambda = 2 \cdot o \cdot c/2^{f}$ where the $c$ is the number of cells on each bucket (typically four as discussed before).  Therefore, the probability that there are $i$ elements with values \{$p_1$,$p_2$\} can be approximated as:

$$
P(i) \cong  \frac{(\lambda)^i}{i!} \cdot e^{-\lambda}
$$

where $\lambda$ will be much smaller than one as for the Cuckoo filter to achieve close to full occupancy we need at least 6 or more bits for the fixed fingerprint. Note that when $i > 8$ we are not able to place the elements in the cuckoo filter and thus the construction of the filter fails. Now, let us consider that a given pair values \{$p_1$,$p_2$\} has $i$ elements with $i \geq 2$, then the probability that for one of those $i$ elements, we have a collision that cannot be removed with the adaptive fingerprint bits can be estimated as: 

\begin{equation}
\left(1- \left(\frac{2^a-1}{2^a}\right)^{i-1} \right)^{2^{\alpha}}
\end{equation}

The exponent $2^{\alpha}$ comes from the number of adaptations that we can do. While the negative term is the probability that the $i-1$ values of the fingerprints of the other $i-1$ elements are all different from the one we are considering.  Thus one minus it is the probability of failing.
Since we have $i$ elements, and assuming that the probability of failure is low, we can approximate the probability of failure for the $i$ elements as: 

\begin{equation}
P_f(i) \cong i \left( 1- \left( \frac{2^a-1}{2^a} \right) ^{i-1} \right) ^{2^{\alpha}}
\end{equation}

This assumes that the probability of failing for one of the elements is independent of the rest which is not true. Therefore, the estimate will not be exact.  In fact for $i = 2$, the failures are totally correlated. Therefore, for that case we will use:

\begin{align}
& P_f(2) \cong \left(1- \left(\frac{2^a-1}{2^a}\right)\right)^{2^{\alpha}} = 
2^{-a2^{\alpha}}
\end{align}

Combining the previous equations, the expected number of collisions $N$ that cannot be removed for the entire table of $m$ buckets can be estimated as:

\begin{align}
& N \cong  \frac{m}{2}  2^{f} \sum_{i = 2}^{8} P(i) P_f(i) = m \cdot 2^{f-1} \cdot \\ \nonumber
& \left(\frac{\lambda^2e^{-\lambda}}{2}2^{-a2^{\alpha}}
+ \sum_{i = 3}^{8} \frac{\lambda^ie^{-\lambda}}{i!} i \cdot \left(1- \left(\frac{2^a-1}{2^a}\right)^{i-1} \right)^{2^{\alpha}} \right)
 \end{align}

Where the summation is truncated at $i = 8$ as for larger values the construction of the cuckoo filter will fail. It should be noted that in the derivation of approximation, it has been assumed that the values of the fingerprints and the positions are uniformly distributed over their range of possible values. This may not be entirely true. For example, if there is a position to which many elements map, the second positions for that bucket have a larger probability of having the fingerprint value that corresponds to that first position.  

The previous analysis has considered the number of collisions that we cannot remove during the initialization of the filter. However, collisions can also occur during its operation as elements are added and removed. Let us consider that the filter has already an occupancy of $o$ and we remove one element and add a new one. Then, the probability $F$ that the new element creates a collision that can not be removed is: 

\begin{align}
 & F \cong \sum_{i = 1}^{7} P(i) \cdot P_f(i+1) = \\ \nonumber
 & \lambda e^{-\lambda} 2^{-a2^{\alpha}} + \sum_{i = 2}^{7} \frac{\lambda^i}{i!} e^{-\lambda} (i+1) \left(1- \left(\frac{2^a-1}{2^a}\right)^{i} \right)^{2^{\alpha}} 
\end{align}
 
The reasoning is as follows: the newly inserted element will go in the pair of values {$p_1 = h_1(x), f_x$} and {$p_2 = h_1(x)\; xor \; h_2(f_x), f_x$}  that has already $i$ elements with probability $P(i)$. When there is at least one element, a collision will be created that needs to be removed. This removal will fail with probability $P_f(i+1)$.  It should be noted that in the above derivation, we assume that the when pair of values had $i$ elements, the collisions could be removed (as otherwise we would have failed before this insertion) and at the same time we use the $P_f(i+1)$ for the general case.  This can be a good approximation since the probability of failing with $i+1$ elements should be much larger than that of failing with $i$ and therefore the first should have little effect on the second.

\section{Evaluation}

Since the goal is to have a data structure that has no collisions, we first use the analytical approximations to show that for some practical configurations, the probability of having non removable collisions will be very low. Then, we validate the analytical approximations by simulation for different values of the probability going as low as it is practical using the available computing resources. 

\begin{figure}[h!]
  \centering
  \includegraphics[scale=0.26]{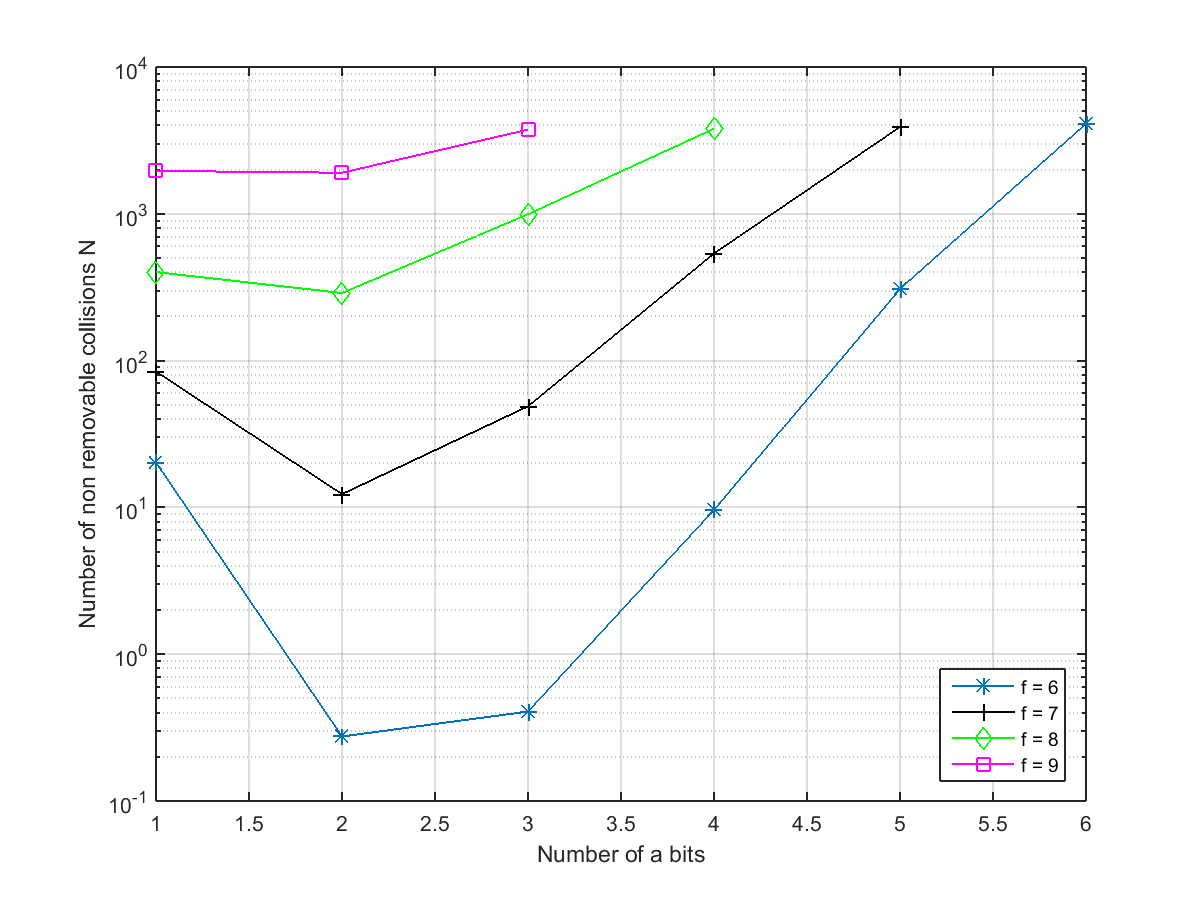}
  \includegraphics[scale=0.26]{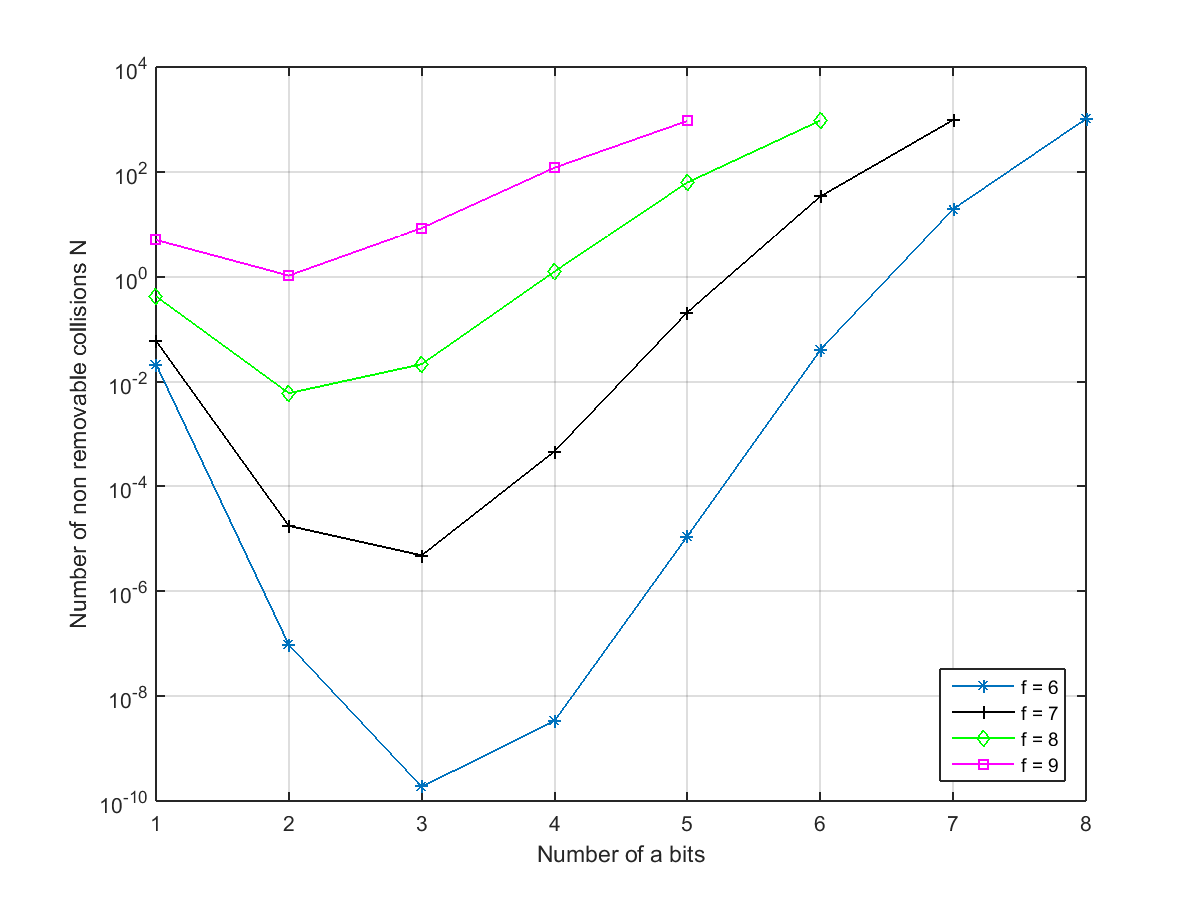}
  \includegraphics[scale=0.26]{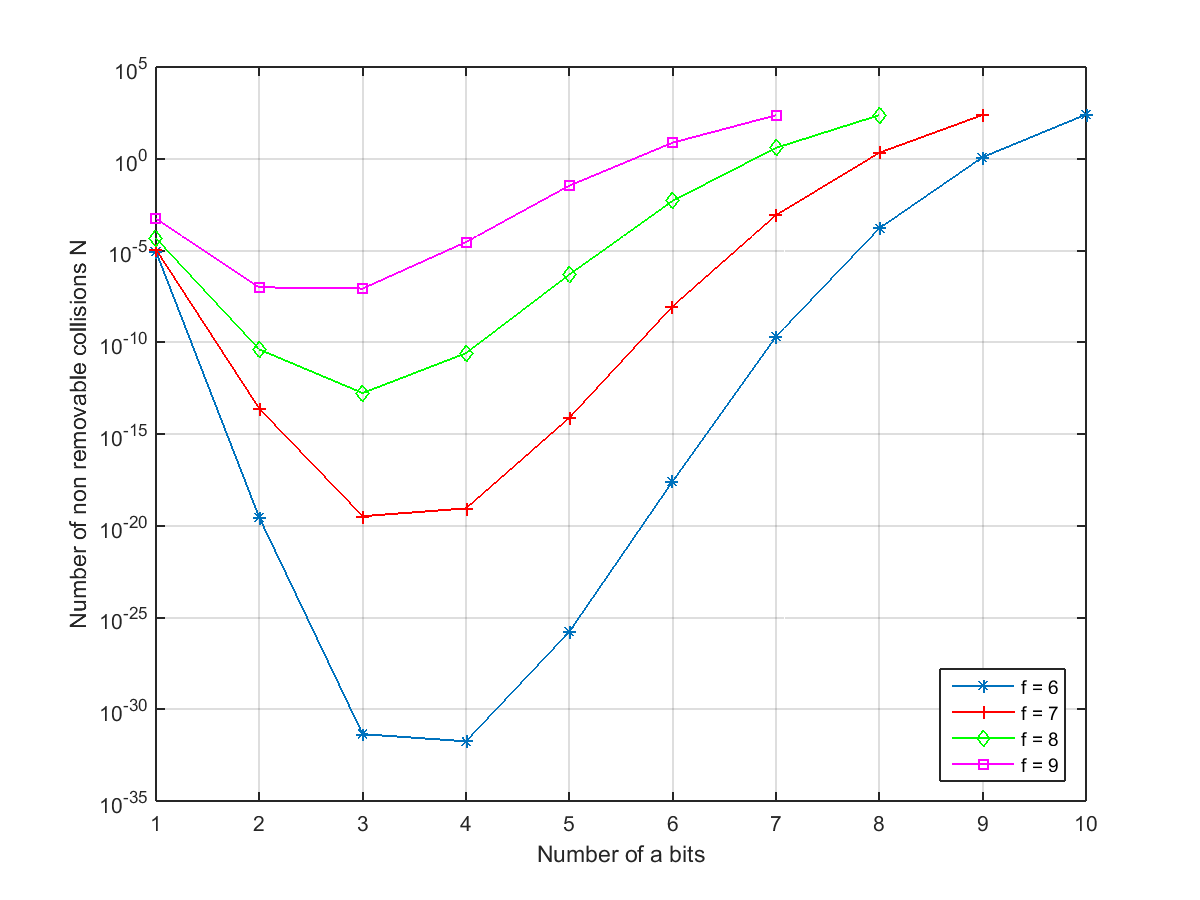}
  
  \caption{Estimated number $N$ of non removable collisions when using 12,14 and 16 fingerprint bits split among $f$,$a$ and $\alpha$ bits according to equation 4.}\label{Fig3}
\end{figure} 

\begin{figure*}[h!]
  \centering
  \includegraphics[scale=0.26]{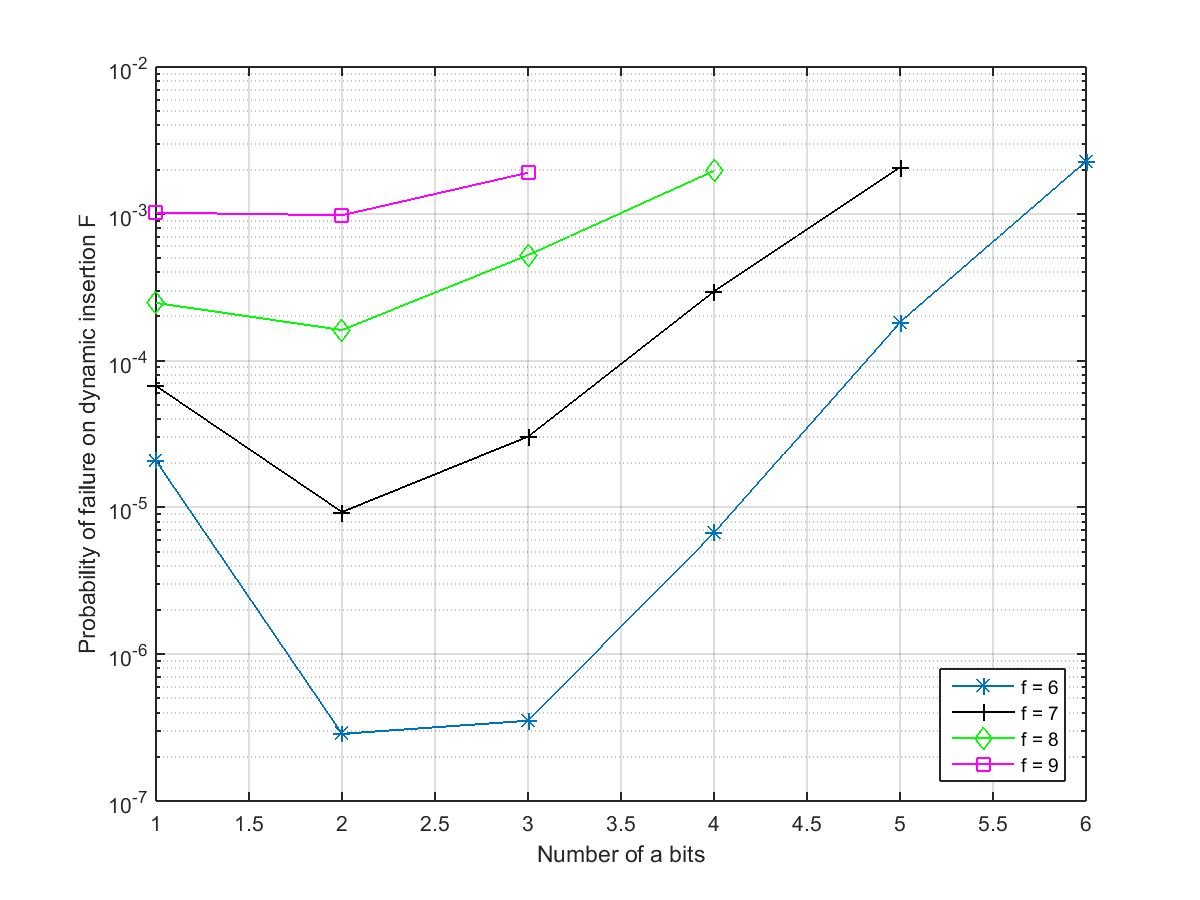}
  \includegraphics[scale=0.26]{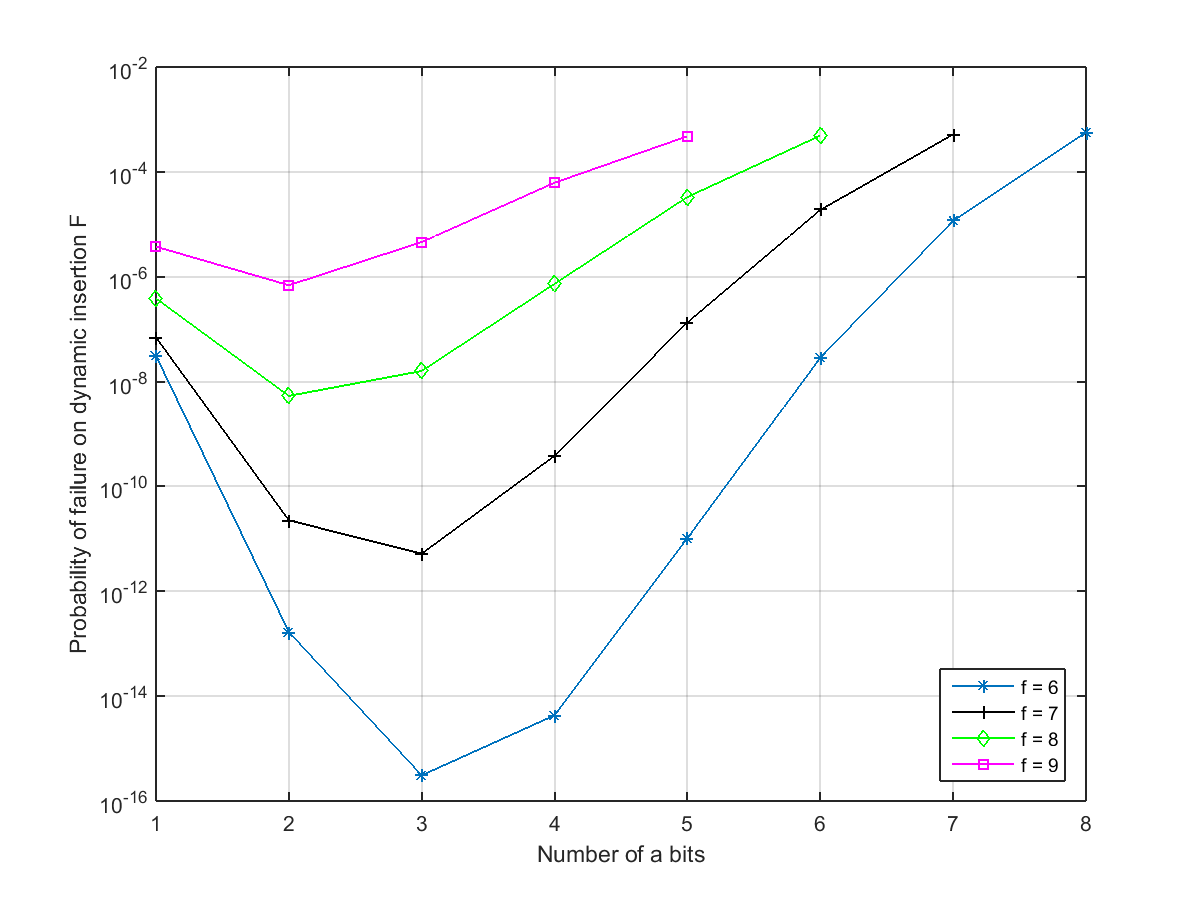}
  \includegraphics[scale=0.26]{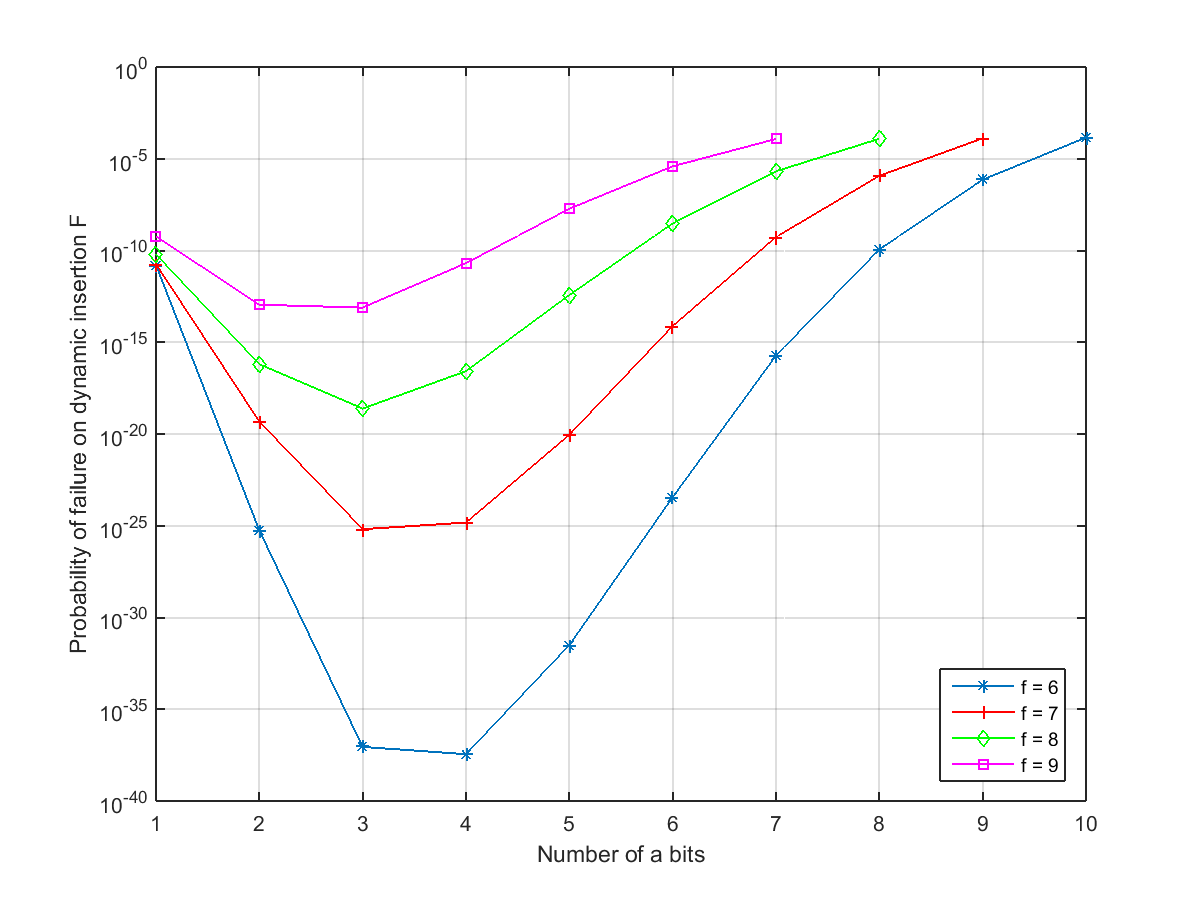}
  \caption{Estimated probability of failure $F$ on dynamic insertion when using 12,14 and 16 fingerprint bits split among $f$,$a$ and $\alpha$ bits  according to equation 5.}\label{Fig4}
\end{figure*} 

To show the potential of CuCoTrack, we have used the approximations presented in the previous section to compute the expected number $N$ of non removable collisions and the probability $F$ of creating a non removable collision when inserting a new element. An occupancy $o$ of 95\% has been assumed that is close to the maximum occupancy achievable by a cuckoo filter \cite{CF}. Three values have been considered for the total number of bits in the fingerprint: 12,14 and 16. Those are split between the fixed part $f$ and the adaptive part $a,\alpha$ assigning at least 6 bits to the fixed part. For the adaptive part, several configurations are tested by changing $a$ and $\alpha$. The results are shown in Figures \ref{Fig3} and \ref{Fig4}. 

It can be seen, that with 16 bits we are able to provide a very small probability when using 6,7 or 8 bits for the fixed fingerprint $f$. If we can use 18 or 20 bits, we can achieve close to zero collisions for larger values of $f$. Therefore, from the analytical approximations, we can conclude that CucoTrack can provide a practically collision free solution when we have 16 or more bits for the fingerprint. Looking at the dynamic part of the fingerprint, it seems that the best split of the bits is approximately in half for $a$ and half for $\alpha$ with a bias towards smaller values of $a$. 
  
To validate the analytical approximations, the proposed scheme has been simulated. This has been done for tables of 512K buckets each having 4 cells giving a total of 4M cells. 

Firstly, two configurations with 12 and 14 bits were tested to check the accuracy of the approximations for $N$ and $F$. In the case of 12 bits, $f$ was fixed to 6 and for 14 bits to 8. Then $a$ was varied from 1 to 5. This corresponds to the blue line in the first plots of Figures \ref{Fig3},\ref{Fig4} and to the green line in the second plots. In these simulations, the filter was constructed 10000 times and for each of the runs, once 95\% occupancy was reached, one million replacement operations were done. The average across all the runs is reported. The comparison of the simulation results with the theoretical estimates showed a worst case deviation of less than 15\%. Therefore it seems that the approximations are reasonably accurate at least to probabilities of dynamic failure down to $10^{-8}$.  



Since the goal of the scheme is to provide close to zero collisions, the next step was to test a configuration with 14 bits that has a lower probability of failure. In particular, $f = 7, a = 4$ and $\alpha = 3$ was tested with one hundred thousand filter constructions. The results again matched the theoretical estimates with less than 15\% error.  These results suggest that the equations obtained in the analysis can be used to estimate the probability of collisions when the probability of failure is even lower (in this case it was approximately $3.8 \cdot 10^{-10}$).

In a final experiment, we have checked that a configuration with a low probability of failure does not suffer non removable collisions. In particular, we have selected 16 bits with $f$ = 8, $a$ = 3 and $\alpha$ = 5.  For this configuration, we have run one hundred thousand filter constructions and for each we have done one million replacements. Non removable collisions were not seen on all the runs. This confirms that CucoTrack can reliably remove collisions when using fingerprints of 16 bits.    

\section{Conclusions} 

In this paper, CuCoTrack, a new data structure to perform connection tracking has been proposed. CuCotrack is based on cuckoo hashing and enables a collision free scheme using small fingerprints instead of storing the complete 5-tuple. This reduces significantly the memory needed to store the connections and the memory bandwidth for queries. The proposed scheme has been analyzed theoretically to show that CuCoTrack is able to significantly reduce the collisions when 12 bits are used for the fingerprint and practically eliminate them when we can afford to use 16 bits or more for the fingerprint. This has been corroborated in simulation for probabilities of failures on insertion down to $10^{-10}$.


%


\section*{Acknowledgment}
Pedro Reviriego would like to acknowledge the support of the excellence network Elastic Networks TEC2015-71932-REDT funded by the Spanish Ministry of Economy and Competitivity. Salvatore Pontarelli is partially supported by the EU Commission in the frame of the Horizon 2020 project 5G-PICTURE (grant \#762057).
 

\end{document}